# Scaling Analysis on Indian Foreign Exchange Market

A. Sarkar and P. Barat*

Variable Energy Cyclotron Centre

1/AF Bidhan Nagar, Kolkata-700 064, India

**Abstract:**

*In this paper we investigate the scaling behavior of the average daily exchange rate returns of the Indian Rupee against four foreign currencies namely US Dollar, Euro, Great Britain Pound and Japanese Yen. Average daily exchange rate return of the Indian Rupee against US Dollar is found to exhibit a persistent scaling behavior and follow Levy stable distribution. On the contrary the average daily exchange rate returns of the other three foreign currencies do not show persistency or antipersistency and follow Gaussian distribution.*

Financial markets are complex dynamical systems with a large number of interacting elements. In physics there is a long tradition of studying the complex systems. Recently physicists got interested in the field of economics and a new subject of study "Econophysics" [1] emerged. The study of a financial market is the most complicated and challenging one due to the complexity of its internal elements, external factors acting on it and the unknown nature of the interactions between the different comprising elements. In the recent years, new and sophisticated methods have been invented and developed in statistical and nonlinear physics to study the dynamical and structural properties of various complex systems. These methods have been successfully applied in the field of quantitative economy [1-3], which gave a chance to look at the economical and financial data from a new perspective. The exchange rates between currencies are particularly interesting category of economic data to study as they dictate the economy of most countries. The time dependence of the exchange rates is usually complex in nature and hence, it is interesting to analyze using the newly developed statistical methods. In this paper we report the study of detailed scaling behavior of the average daily exchange rate returns of Indian Rupee (INR) versus four important foreign currencies in Indian economy, namely the US Dollar (USD), the EURO, the Great Britain Pound (GBP) and the Japanese YEN for the past few years. India, being the country with second largest population in the world, is an important business market for the multinational companies. Therefore the study of the average daily exchange rate returns of Indian Rupee with respect to the four foreign currencies is very significant and relevant from the economic point of view.

* Corresponding author:, Variable Energy Cyclotron Centre, 1/AF Bidhan Nagar, Kolkata 700064, India, Phone: +91-33-23371230, Fax: +91-33-23346871, e-mail: pbarat@veccal.ernet.in

Scaling as a manifestation of underlying dynamics is familiar throughout physics. It has been instrumental in helping scientists gain deeper insights into problems ranging across the entire spectrum of science and technology. Scaling laws typically reflect underlying generic features and physical principles that are independent of detailed dynamics or characteristics of particular models. Scale invariance seems to be widespread in natural systems [4]. Numerous examples of scale invariance properties can be found in the literature like earthquakes, clouds, networks etc. [5-8]. Scaling investigation in the financial data has been recently got much importance. In the literature, many empirical studies can be found which show that financial time series exhibit scaling like characteristics [9-13]. However, some literature continued to question the evidence of the scaling laws in the foreign exchange markets. LeBaron [14,15] examined the theoretical foundation of scaling laws and demonstrated that many graphical scaling results could have been generated by a simple stochastic volatility model. He suggested that the dependence in the financial time series might be the key cause in the apparent scaling observed. His model was able to produce visual power-laws and long memory similar to those observed in financial data of comparable sample sizes. However Stanley *et al.* [16] pointed out that a three-factor model cannot generate power-law behavior.

Thus, it is still an open question of the scaling behavior of financial time series. Recently, Matia *et al*. [17] have carried out analysis on the 49 largest stocks of the National Stock Exchange of India. They have shown that the stock price fluctuations in India are scale dependent. In this work we have studied the daily evolution of the currency exchange data [18] of INR-USD, INR-EURO, INR-GBP and INR-YEN for the period of 25th August 1998 to 31st August 2004 (for INR-EURO the time period is 1st January 1999 to 31st August 2004) using two newly developed methods namely (i) the Finite Variance Scaling Method (FVSM) (ii) the Diffusion Entropy Analysis (DEA) to reveal the exact scaling behavior of the average daily exchange rate returns. The return Z(t) of the exchange rate time series X(t) is defined as $Z(t) = \ln\frac{X(t+1)}{X(t)}$. Fig. 1 (a) shows the variation of the average daily exchange rates of INR against USD. The variations of the daily exchange rate returns of INR against USD, EURO, GBP and YEN are shown in Fig 1 (b), (c), (d) and (e) respectively.

Two complementary scaling analysis methods: FVSM and DEA [19-21] together are found to be very efficient to detect the exact scaling behavior of complex dynamical systems. The need for using these two methods to analyze the scaling properties of a time series is to discriminate the stochastic nature of the data: Gaussian or Levy [21,22]. These methods are based on the prescription that numbers in a time series $\{Z(t_i)\}$ are the fluctuations of a diffusion trajectory; see Refs. [20,23,24] for details. Therefore, we shift our attention from the time series $\{Z(t_i)\}$ to probability density function (pdf) $p(x,t)$ of the corresponding diffusion process. Here $x$ denotes the variable collecting the fluctuations and is referred to as the diffusion variable. The scaling property of $p(x,t)$ takes the form

$$p(x,t) = \frac{1}{t^{\delta}} F\left(\frac{x}{t^{\delta}}\right) \qquad (1)$$

In the FVSM one examines the scaling properties of the second moment of the diffusion process generated by a time series. One version of FVSM is the standard deviation analysis (SDA) [19], which is based on the evaluation of the standard deviation $D(t)$ of the variable $x$, and yields [4,19].

$$D(t) = \sqrt{\langle x^2;t\rangle - \langle x;t\rangle^2} \propto t^{\gamma} \qquad (2)$$

The exponent $\gamma$ is interpreted as the scaling exponent.

DEA introduced recently by Scafetta *et al*. [19] focuses on the scaling exponent δ evaluated through the Shannon entropy *s(t)* of the diffusion generated by the fluctuations $\{Z(t_i)\}$ of the time series using the pdf (1) [19,20]. Here, the pdf of the diffusion process, $p(x,t)$, is evaluated by means of the subtrajectories $x(t_n) = \sum_{i=0}^{n} Z(t_{i+n})$ with $n$ =0,1,.. Using Eq. (1) we arrive at the expression for *s(t)* as

$$s(t) = -A + \delta \ln(t) \qquad (A = \text{Constant}) \qquad (3)$$

Eq. (3) indicates that in the case of a diffusion process with a scaling pdf, its entropy $s(t)$ increases linearly with $\ln(t)$. Finally we compare $\gamma$ and $\delta$. For fractional Brownian motion the scaling exponent $\delta$ coincides with the $\gamma$ [20]. For random noise with finite variance, the pdf $p(x,t)$ will converge to a Gaussian distribution with $\gamma = \delta =0.5$. If $\gamma \neq \delta$ the scaling represents anomalous behavior.

The plots of SDA and DEA for the average daily exchange rate returns of the four foreign currencies are shown in Fig. 2 and Fig. 3 respectively. The scaling exponents obtained from the plots of SDA and DEA are listed in Table I. The values of $\gamma$ and $\delta$ clearly reflect that the INR-USD exchange rate returns behave in a different manner with respect to the other three exchange rate returns. For INR-USD exchange rate returns the scaling exponents are found to be greater than 0.5 indicating a persistent scaling behavior. While the unequal values of $\gamma$ and $\delta$ implies anomalous scaling. For the other three exchange rate returns, the values of $\gamma$ and $\delta$ are almost equal to 0.5 within their statistical error limit, signifying absence of persistency and antipersistency in those cases. The results obtained from SDA and DEA seem to be surprising as all the exchange rate return time series data are from the same foreign exchange market. To confirm the observations obtained from the results of SDA and DEA, we applied another well-established method namely R/S Analysis to the average daily exchange rate return data.

Range/Standard (R/S) deviation analysis also referred to as rescaled-range analysis was originally developed by Hurst [25]. The *R*/*S* analysis is performed on the discrete time-series data set $\{Z(t_i)\}$ of dimension $N$ by calculating the accumulated departure, $Y(n,N)$, according to the following formula:

$$Y(n,N) = \sum_{i=1}^{n} (Z(t_i) - \bar{Z}(N)) ...........0 < n \leq N \quad (4)$$

Where $\bar{Z}(N)$ is the mean value of $\{Z(t_i)\}$. The range of the $Y(n,N)$, is given by

$$R(N) = \max\{Y(n,N)\} - \min\{Y(n,N)\} \quad (5)$$

Finally, the rescaled-range (*R(N)*/*S(N)*) is determined as a function of N, where S(N) is the standard deviation of $\{Z(t_i)\}$. Scaling in this case implies

$$R(N)/S(N) \propto N^H \quad (6)$$

Where $H$ is called the Hurst exponent. $H = 0.5$ implies statistical independence and ordinary Brownian motion. $H > 0.5$ and $H < 0.5$ respectively imply persistent and antipersistent long range correlation. The plots and the Hurst exponents obtained from the R/S analysis for the average daily exchange rate returns of the four foreign currencies are shown in Fig.4. The results of the R/S analysis confirm the persistent scaling in INR-USD exchange rate return data and randomness in other exchange rate returns.

The primary objectives of these analyses were to find the generic feature of these time series data, their long range correlation and their robustness to retain the scaling property. To verify the robustness of the observed scaling property of INR-USD exchange rate return data, we corrupted 2% of the exchange rate return data at random locations by adding noise of magnitude of the multiple of the standard deviation (std).

We found that addition of noise of magnitude of five times of the std the scaling exponents did not change and the scaling behavior is retained by an addition of noise of magnitude of fifteen times of std. Which confirms the robustness of the scaling property of the INR-USD exchange rate return data. We have also analyzed the probability density distribution of the exchange rate returns. The distributions are fitted with Levy stable distribution, which is expressed in terms of its Fourier transform or characteristic function, $\varphi(q)$, where $q$ is the Fourier transformed variable. The general form of the characteristic function of a Levy stable distribution is:

$$\ln \varphi(q) = i\xi q - \eta |q|^{\alpha} \left[1 + i\beta \frac{q}{|q|} \tan\left(\frac{\pi\alpha}{2}\right)\right] \quad \text{for } [\alpha \neq 1] \tag{7}$$

$$= i\xi q - \eta |q| \left[1 + i\beta \frac{q}{|q|} \frac{2}{\pi} \ln |q|\right] \quad \text{for } [\alpha = 1]$$

where $\alpha \in (0,2]$ is an index of stability also called the tail index, $\beta \in [-1,1]$ is a skewness or asymmetry parameter, $\eta > 0$ is a scale parameter, and $\xi \in \mathbb{R}$ is a location parameter which is also called mean. For Cauchy and Gaussian distribution, the values of $\alpha$ are equal to 1 and 2 respectively. The fits [26] of the Levy stable distribution for the four exchange rate returns are shown in Fig. 5. Insets in the figures show the plots in log-log scale. The parameters of the fitted Levy stable distribution for the average daily exchange rate returns of the four currencies are presented in Table II. From Table II it is seen that the value of $\alpha$ in case of INR-USD exchange rate is 1.3307 indicating the distribution is of Levy type but for the other cases $\alpha$ values are close to the Gaussian limit 2. Which is also an indication of the randomness in those exchange rate returns.

The political development inside and outside a country affects its economy and the foreign exchange market. The cross currency volatility also influences a particular kind of exchange rate. The world economy experienced one of the worst shocks in the aftermath of September 11, 2001 events in the United States. Foreign exchange market in India also became volatile (shown in Fig. 1a). Another large fluctuation in INR-USD exchange rate is observed around 31 March 2004. These fluctuations in the INR-USD exchange rate did not affect its robust scaling property. We argue this is due to the dissipation of the fluctuation in the vast economy of a country like India. The interacting elements provide a retarding path to the fluctuations in a financial market. As the number of interacting element increases the channel for the fluctuation dissipation gets broaden. USD is the most important foreign currency in the Indian economy. Hence, the number of interacting elements is more in the INR-USD exchange market. Possibly this is the reason behind the observed robustness of the scaling property in the INR-USD average daily exchange rate returns.

The exchange rate management policy continues its focus on smoothing excessive volatility in the exchange rate with no fixed rate target, while allowing the underlying demand and supply conditions to determine the exchange rate movements over a period in an orderly way. Towards this end, the scaling analysis of the foreign exchange rate data is of prime importance. We have carried out extensive studies on the average daily exchange rate returns from Indian foreign exchange market. From the analyses we have found that the average daily exchange rate return of USD exhibits scaling and follows Levy Stable distribution. On the contrary, the average daily exchange rate returns of the other foreign currencies namely EURO, GBP and YEN do not follow persistency or antipersistency and they are found to obey Gaussian distribution.

**Figure Captions:**

Fig. 1. (a)Variation of the average daily INR-USD exchange rate. Variation of the return of the average daily exchange rates of (b) INR-USD (c) INR-EURO (d) INR-GBP and (e) INR-YEN.

Fig. 2. SDA of the average daily exchange rate returns.

Fig. 3. DEA of the average daily exchange rate returns.

Fig. 4. R/S analysis of the average daily exchange rate returns.

Fig. 5. Levy stable distribution fit of the (a) INR-USD (b) INR-EURO (c) INR-GBP and (d) INR-YEN average daily exchange rate return distributions. Insets in the figures show the plots in log-log scale (the return axis is shifted by 2 to show the negative tail).

TABLE I. Scaling exponents $\gamma$ and $\delta$ obtained from SDA and DEA respectively for the average daily exchange rate returns of Indian Rupee versus the four foreign currencies.

| Data | Method of Analysis | |
|---|---|---|
| | SDA ($\gamma$) | DEA ($\delta$) |
| USD | 0.59(±0.02) | 0.64(±0.02) |
| EURO | 0.50(±0.02) | 0.49(±0.02) |
| GBP | 0.48(±0.02) | 0.49(±0.02) |
| YEN | 0.51(±0.02) | 0.48(±0.02) |

Table II. Parameters of the Levy stable distribution fit for the average daily exchange rate returns of the four currencies.

| Data | $\alpha$ | $\beta$ | $\eta$ | $\xi$ |
|---|---|---|---|---|
| USD | 1.3307 | 0.1631 | $0.5376\times10^{-3}$ | $-0.4517\times10^{-4}$ |
| EURO | 1.9900 | -0.9997 | $0.5117\times10^{-2}$ | $0.2436\times10^{-3}$ |
| GBP | 1.8860 | -0.1005 | $0.3594\times10^{-2}$ | $0.1768\times10^{-3}$ |
| YEN | 1.8555 | 0.0713 | $0.4357\times10^{-2}$ | $0.2889\times10^{-4}$ |

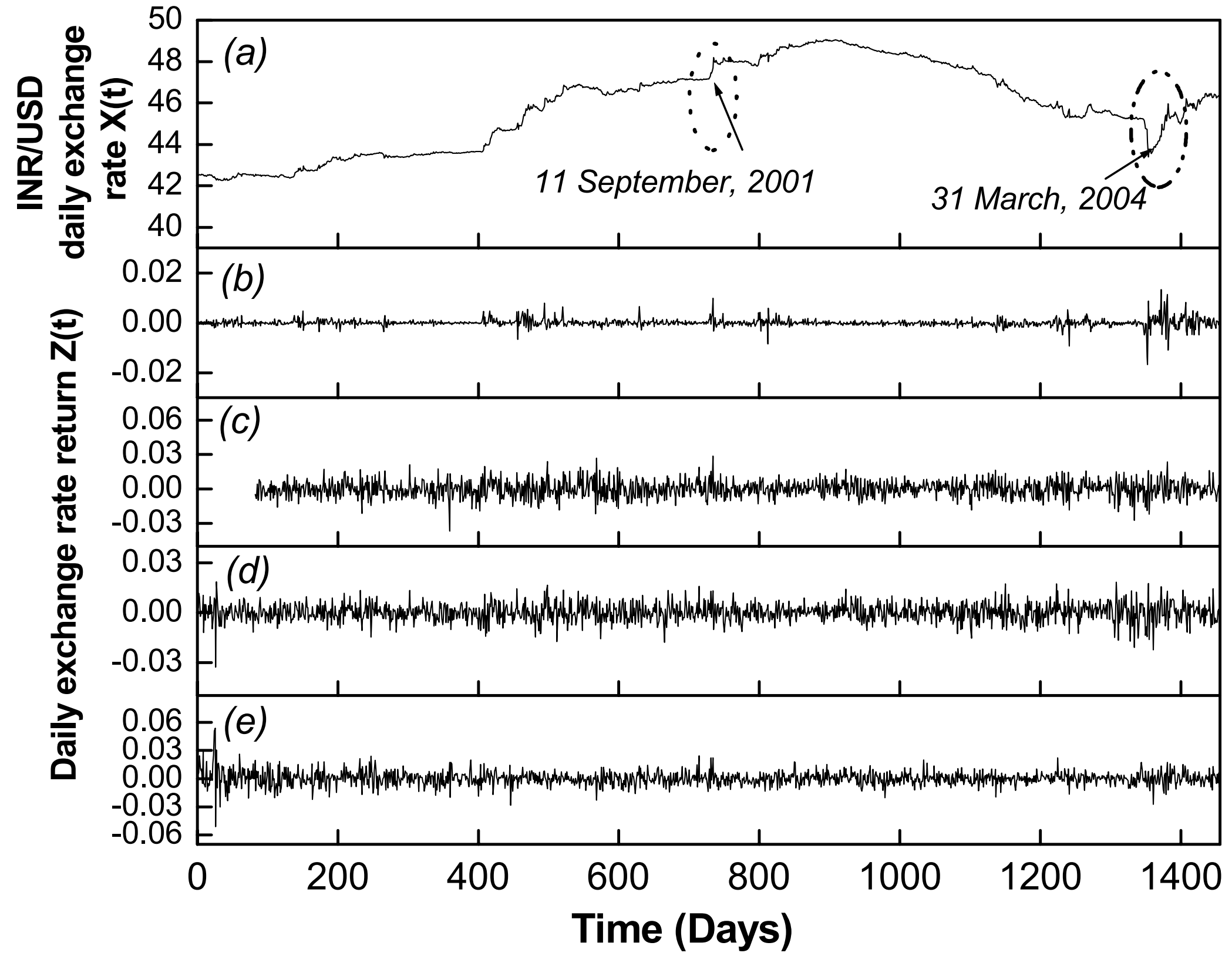

FIG. 1

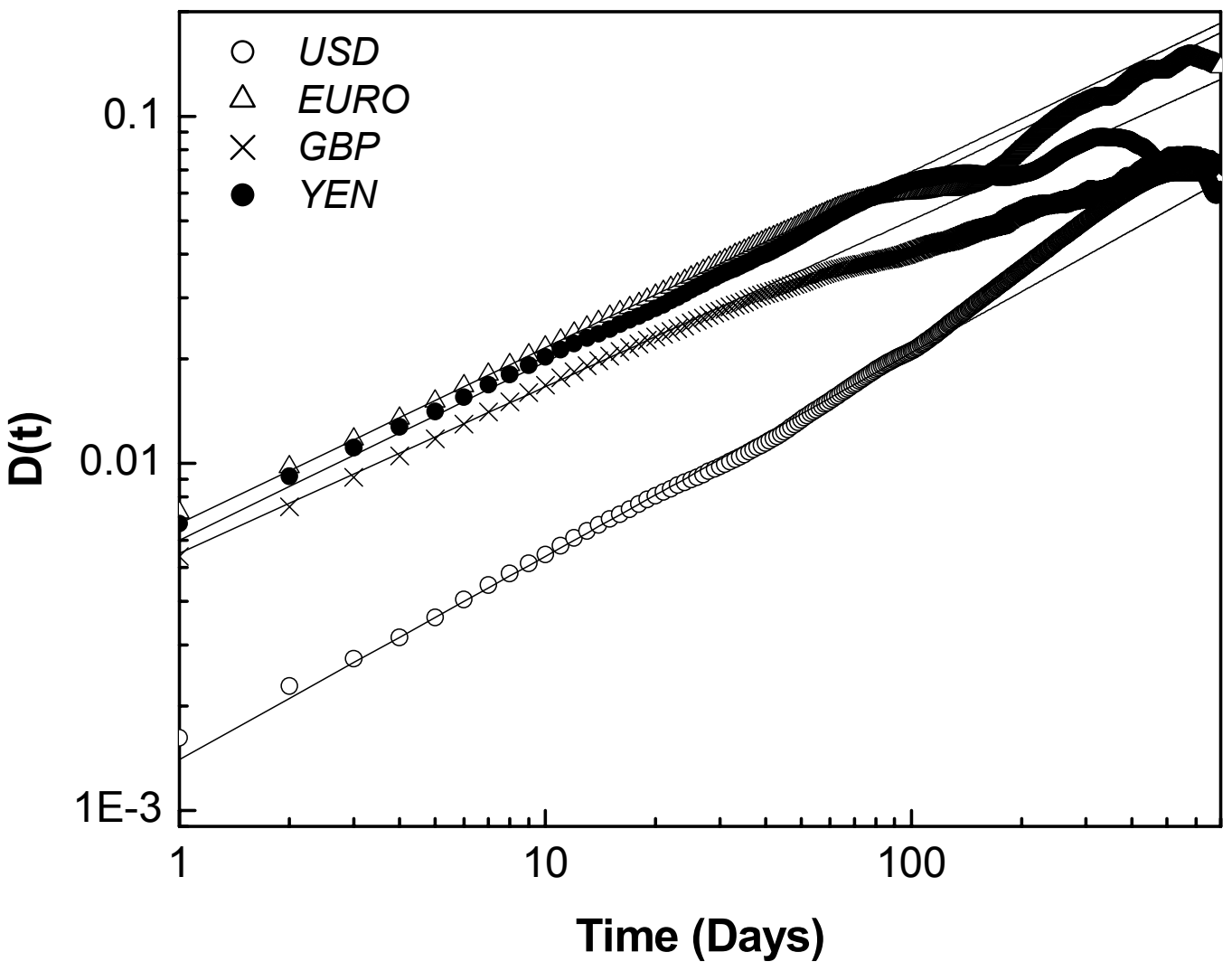

FIG. 2

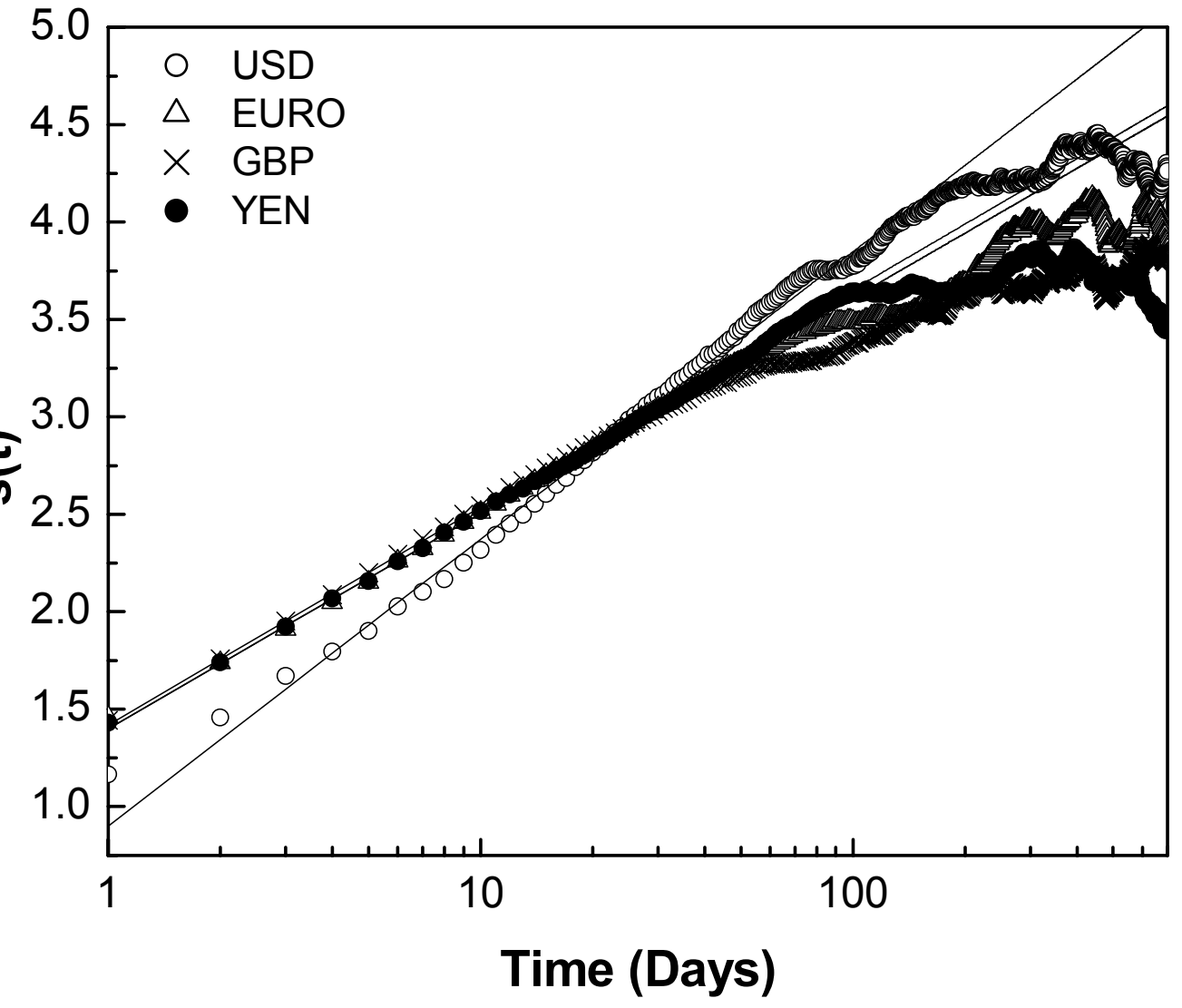

FIG. 3

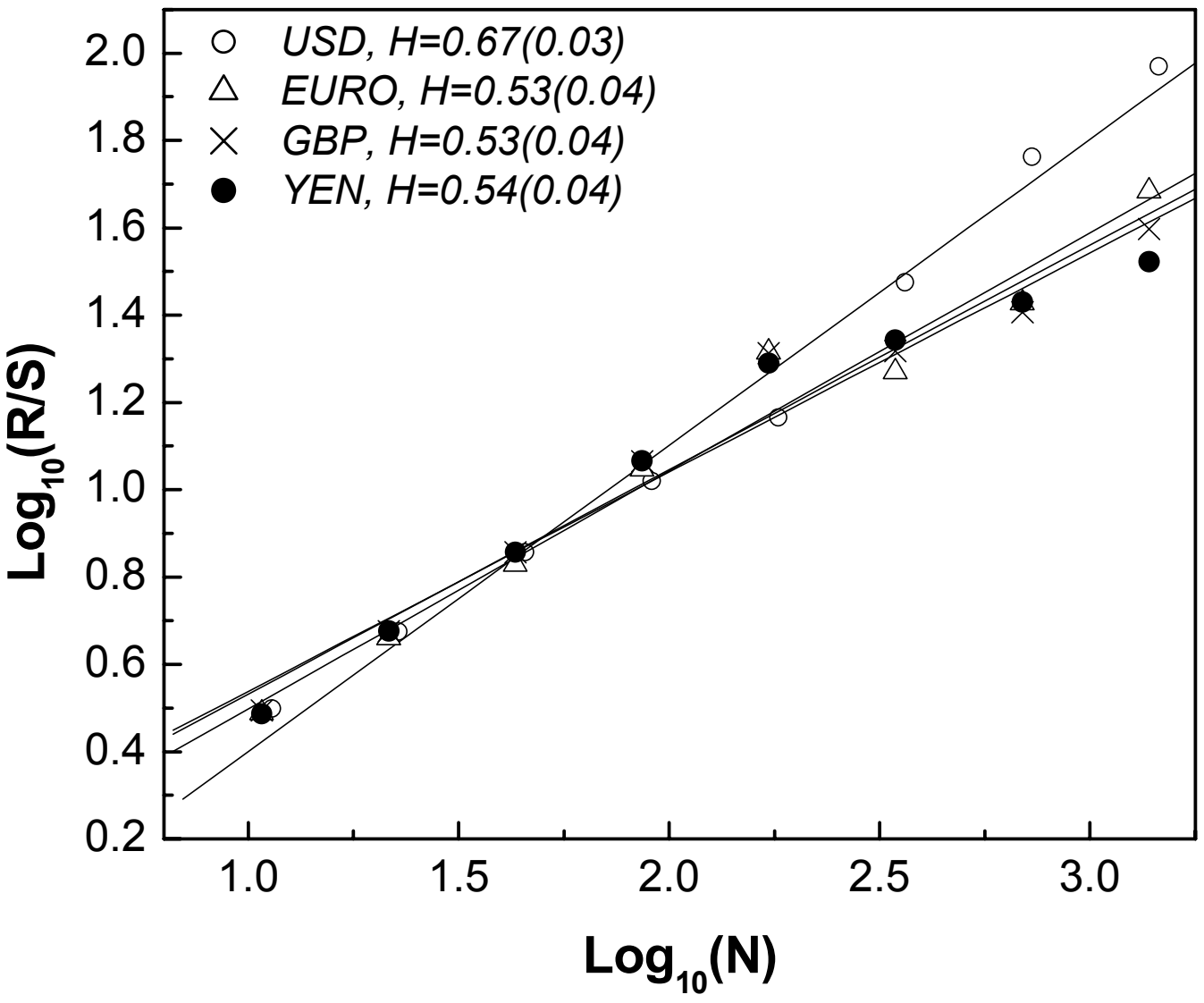

FIG. 4

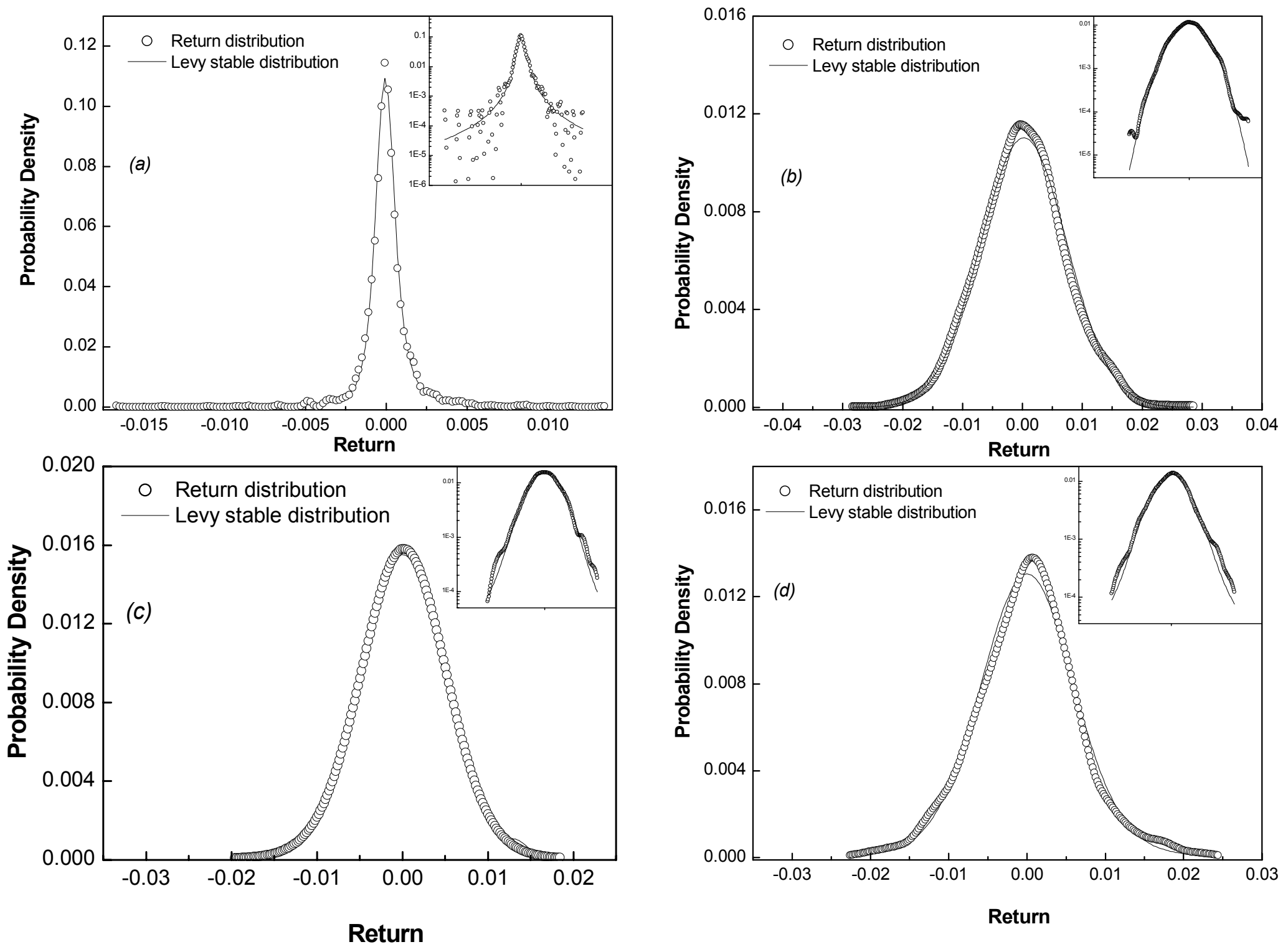

FIG. 5